\documentclass[12pt,preprint]{aastex}
\begin{document}
\title{Stellar Activity and the Str\"{o}mgren Photometric Metallicity 
Calibration of Intermediate-Type Dwarf Stars}

\bigskip
\author{Sarah L. Martell and Graeme H. Smith} 
\affil{University of California Observatories/Lick Observatory, University of 
California}
\affil{Santa Cruz, California 95064}
\bigskip
\begin{abstract} 
\noindent
We consider the effect of stellar activity, as measured by X-ray 
luminosity, on metallicities of Solar-neighborhood F and G dwarfs derived 
from Str\"{o}mgren photometry.  Rocha-Pinto \& Maciel found evidence 
that Str\"{o}mgren colors systematically underpredict [Fe/H] for stars
with extremely high \ion{Ca}{2} H \& K emission.  We investigate
whether a recent photometric metallicity calibration derived by
Martell \& Laughlin might be subject to this effect, and whether the
amount of underprediction could reliably be expressed as a function of 
$\log (L_{\rm X}/L_{\rm bol})$.  Among those calibration stars used by
Martell \& Laughlin which are also in the Bright Star Catalogue and
detected in the \textit{ROSAT} All-Sky Survey there is no evidence for
a correlation between photometric metallicity and stellar activity.
However, many of the ``very active stars'' on which the Rocha-Pinto \&
Maciel result was based are members of interacting binaries or are
very young in age, and are not included in the X-ray sample that we
are using.  Among normal dwarf stars it appears that stellar activity
has little effect on the metallicity calibration of Str\"{o}mgren colors.

\end{abstract}


\section{Introduction}

The effect of chromospheric activity on photometric techniques for measuring 
the metallicities of stars has been explored by several authors commencing 
with Giampapa, Worden, \& Gilliam (1979). With recent 
large photometric and spectroscopic surveys the question can be addressed 
in a statistically meaningful way (see e.g., West et al. 2004).  
Rocha-Pinto \& Maciel (1998) correlated the calcium emission line 
index $\log R_{HK}^{\prime}$ against metallicities calculated from 
Str\"{o}mgren photometry, and found that their most-active stars had 
surprisingly low values of inferred ${\rm [Fe/H]_{phot}}$. 

In highly-active stars, the equivalent width of metallic absorption lines 
can be reduced by chromospheric emission in the lines (see e.g., Basri 
et al. 1989). For extremely-active stars this effect may reduce the
Str\"{o}mgren $m_1$ index, which is intended to measure line
blanketing (Crawford 1975), leading to a falsely low photometric
metallicity.  Rocha-Pinto \& Maciel (1998) appealed to an
activity-$m_1$ correlation to explain the apparent low photometric
metallicity (and corresponding apparent old age) of the most-active
nearby stars as an artifact of their chromospheric activity.  In this
paper we make a similar comparison, adopting as an indicator of
stellar activity the soft X-ray luminosity measured by the
\textit{ROSAT} satellite, and using a photometric metallicity
calibration (Martell \& Laughlin 2002, hereafter ML02) which was
developed to be more accurate for higher-metallicity stars than the
Schuster \& Nissen (1989, hereafter SN89) calibration.

\section{The metallicity data set}

In ML02 the authors used a set of 664 F, G, and K dwarfs located within 
100 pc of the Sun to derive an empirical relation between Str\"{o}mgren 
photometric indices and metallicity.  
The selection criteria for the ML02 ``calibration stars'' are as follows:
they are the members of the Cayrel de Strobel, Soubiran,  
\&  Ralite (2001) compilation of [Fe/H] abundances which have absolute
magnitudes $M_V > +1.0$, \textit{Hipparcos} parallaxes 
greater than 0.01\arcsec, and which are also in the Hauck-Mermilliod (1998) 
compilation of Str\"{o}mgren photometry.  The calibration used a 
Levenberg-Marquardt algorithm (e.g., Press et al. 1992) to find the 
coefficients for a general third-order polynomial relating [Fe/H] to 
the Str\"{o}mgren indices $(b-y)$, $m_1$, and $c_1$.  When the
distributions of the residuals (i.e., ${\rm [Fe/H]_{spec} -
  [Fe/H]_{phot}}$) for the ML02 and SN89 calibrations were compared,
the former was found to be more accurate. This can be seen in
Figure 1, which shows a Gaussian fit to each distribution. Both the
central offset and the half-width at half-maximum of the fits are
smaller for the ML02 calibration.  

We have modified slightly the methodology of ML02: in the Cayrel de 
Strobel et al. (2001) metallicity catalog, many stars have multiple [Fe/H] 
measurements, which were treated as separate objects for the purposes of 
the ML02 polynomial fitting.  For the present work we 
averaged such multiple [Fe/H] values together, both to reduce the effect of 
outlying measurements, and to prevent multiple-counting of stars in 
our histograms and plots.  For stars with multiple observations, we
took the observational error in [Fe/H] to be the standard deviation in
the mean, calculated from those multiple measurements.  For stars with
single observations, or where the standard deviation was zero, we
adopted as the error the mean of the standard deviation for all of the
multiply-observed stars.  That quantity has a value of 0.0824 dex.  We
then refitted the ML02 Str\"{o}mgren photometry-metallicity relation,
and while the values of the coefficients did change, the overall
quality of the fit stayed roughly constant.  
The same residual-distribution test was done as in ML02,
and the results are also shown in Figure 1. The Gaussian fits to the
new residual distribution (hereafter MS04) and that of ML02 are almost
equivalent: the center falls at $-0.0248$ for the ML02 calibration,
and at $-0.0266$ for the MS04 calibration.  The HWHM for ML02 is
$0.0868$, and $0.0890$ for MS04.  In the residual distribution for the
SN89 calibration, the center of the Gaussian fit is at $-0.0517$, and
the HWHM is $0.1097$, values which are clearly different from the ML02
and MS04 calibrations.
 
 The resulting calibration is 

\begin{eqnarray}
[{\rm Fe/H}]_{\rm phot} = &-&41.836891+153.92203(b-y)+53.678346m_1+129.01008c_1 \nonumber \\
&-&101.47843(b-y)^2+161.87500m_1^2-150.07528c_1^2-412.75949(b-y)m_1 
\nonumber \\
&-&370.84617(b-y)c_1+52.187608m_1c_1-103.14707(b-y)^3+81.084037m_1^3 
\nonumber \\
&+&53.244338c_1^3+651.10576(b-y)^2m_1+204.52658(b-y)^2c_1 \nonumber \\
&-&452.44692m_1^2(b-y)-80.536525m_1^2c_1+247.37448c_1^2(b-y) \nonumber \\
&-&90.169531c_1^2m_1+128.07586(b-y)m_1c_1 \nonumber \\
\end{eqnarray}

The differences between photometric metallicities derived from this new 
calibration and those of ML02 and Schuster \& Nissen (1989) are shown in 
Figures 2 and 3 respectively as a function of the spectroscopic
metallicity of the calibrating stars.  These diagrams give an
appreciation for the uncertainty in using these fitting functions to
derive [Fe/H] from Str\"{o}mgren colors.  We should point out that the
calibration stars for ML02 and MS04 do not extend to metallicities as
low as the calibrators used by Schuster \& Nissen (1989), and these
former calibrations should only be employed over the range in [Fe/H]
shown in Figures 2 and 3.

The change in coefficients between the ML02 and MS04 calibrations lead
us to investigate how strongly the coefficients depend on the assumed
errors in the spectroscopic [Fe/H] values of the calibrating stars.
If the errors are assumed to be the same for all stars, the
coefficients are insensitive to the value of the assumed error: they
are the same whether the assumed error is 0.05 dex, 0.10 dex, or even
0.20 dex.    
 
By contrast, more complicated behavior resulted when we allowed the errors 
in [Fe/H] to vary among the calibration stars.  For each star with multiple 
[Fe/H] measurements we calculated the mean [Fe/H] value, the standard
deviation $\sigma$ in these [Fe/H] values, and the standard deviation
in the mean $\sigma_m$.  We then ran two different fits for the
calibration stars, assuming in both cases that the error in [Fe/H] for
each multiply-observed star was equal to the individual value of
$\sigma_m$ calculated for that star.  These two fits differed in the
error adopted for all of the singly-observed stars; in the first case
this error was taken to be the average value of $\sigma$ from the
multiply-observed stars, while in the second case an error twice this
amount was adopted. It is the former of these fits that corresponds to
equation (1).  The coefficients, as a rule, were larger for the 
second fit. Most of the terms involving $(b-y)$ and $m_{1}$ stayed fairly
constant, although two of the largest changes were in the coefficients
of $(b-y)^{2}m_{1}$ and $(b-y)m_{1}^{2}$, which are the two largest
coefficients in the calibration. 

We also experimented with setting to zero the four coefficients whose
values varied the most with changes in the errors.  This caused the
other coefficients to all decrease.  However, the quality of the fits,
as measured by Gaussian parameters, stayed roughly constant as we
varied the assumed values of the errors, and all had centers closer to
zero and smaller HWHMs than the SN89 calibration.

\section{The ROSAT data set}

The \textit{ROSAT} observatory conducted an All-Sky Survey (RASS) of X-ray 
sources (Voges et al. 1999) during 1990 and 1991 using the onboard
PSPC imaging detector (Pfeffermann et al. 1987). H\"{u}nsch, Schmitt, \& 
Voges (1998) searched the RASS data for detections at the locations of all 
main sequence and subgiant stars listed in the Bright Star Catalogue
(BSC; Hoffleit \& Warren 1991).
We sifted the ``calibration stars'' used by ML02 
for those also included in the catalog of H\"{u}nsch, Schmitt, \& Voges 
(1998). This produced a set of 146 stars that 
we refer to as the ``RASS-BSC-calibration'' sample.
Figures 4 and 5 show the distribution of distances and spectroscopic 
metallicities respectively for these stars.
Distances derived from the \textit{Hipparcos} Catalogue (ESA 1997),
together with a table of bolometric corrections from 
Allen's Astrophysical Quantities (Cox 2000),
were used to convert the RASS fluxes in the 0.1 - 2.4 keV energy range
into X-ray luminosities and to calculate $\log (L_{\rm X}/L_{\rm bol})$.

According to H\"{u}nsch et al. (1998), the RASS has a typical flux limit of 
$10^{-13}$ erg cm$^{-2}$ s$^{-1}$.
This can readily be seen in Figure 6, which shows a plot of the X-ray 
luminosities of the RASS-BSC-calibration stars versus distance. 
Very few of these stars are more than 60 pc distant, and the 
majority are within 25 pc of the Sun. We note for comparison with 
Figure 6 that a solar-like dwarf with $M_V = 4.7$ at a distance of 20 pc
will have an apparent magnitude of $V = 6.2$, at the limit of the
Bright Star Catalogue. A star at this distance would require an X-ray 
luminosity of $L_{\rm X} > 5 \times 10^{27}$ ergs s$^{-1}$ to be detected
in the RASS. Since the goal of this paper is to investigate whether the
Str\"{o}mgren metallicity calibration is compromised among
stars with high $\log (L_{\rm X}/L_{\rm bol})$,
the RASS flux limit may not adversely bias our analysis: 
the most-active of our calibration 
stars are above the RASS flux limit all the way 
out to 100 pc. In the next section, we test for possible bias by 
showing that the photometric trends for 
the RASS-BSC-calibration stars within 
25 pc are the same as for the full sample.
However, we refrain from using the calibration data to form conclusions
about any possible dependency of X-ray activity on stellar metallicity: the 
RASS flux limit biases against the presence of low-metallicity
stars in our sample because of their relatively low space density in the
Solar neighborhood compared to near-solar abundance stars.
Indeed, nearly all of the RASS-BSC stars found in our
Str\"{o}mgren-[Fe/H] calibration sample have metallicities of 
[Fe/H] $> -0.5$.

\section{ Discussion }

To investigate the question of whether chromospheric activity affects
the metallicity derived from Str\"{o}mgren photometry, we looked for
trends involving $[{\rm Fe/H]_{phot}}$, $[{\rm Fe/H]_{spec}}$, and
$\log (L_{\rm X}/L_{\rm bol})$ among the RASS-BSC-calibration stars.
Figure 7 shows the difference between $[{\rm Fe/H]_{spec}}$ and $[{\rm
    Fe/H]_{phot}}$  derived from equation (1) as a function of $\log
(L_{\rm X}/L_{\rm bol})$ for the full set of RASS-BSC-calibration
stars out to 100 pc, and a best-fit line obtained by a regression of
$[{\rm Fe/H]_{spec}}$ -- $[{\rm Fe/H]_{phot}}$versus $\log (L_{\rm
  X}/L_{\rm bol})$.  The slope of the best-fit line is
0.0121$\pm$0.0144, which is fairly consistent with there being no
trend.  The linear Pearson correlation coefficient calculated for the
data is 0.0696.  Figure 8 shows the same quantities for those
RASS-BSC-calibration stars within 25 pc, and the slope of that
best-fit line is 0.0210$\pm$0.0179.  The correlation coefficient for
these data is 0.1287.  Both correlation coefficients are quite small;
there appears to be no significant evidence that metallicities derived
from equation (1) are compromised by stellar activity among normal solar 
neighborhood dwarf stars.

We have conducted a similar analysis using the same set of stars but the 
photometric metallicity calibration of Schuster \&  Nissen (1989) rather
than equation (1). The results are shown in Figures 9 and 10, again for stars 
within 100 pc and 25 pc of the Sun respectively. Once again there is no 
evidence that the photometric metallicities depart from the spectroscopic
values in any way correlated with stellar activity.  The correlation 
coefficient is 0.0915 for the 100-pc set and 0.1775 for the 25-pc set.

The lack of trends in Figures 7-10 is not surprising: the  
H\"{u}nsch et al. (1998) stars in our calibration set have little overlap with
the Rocha-Pinto \& Maciel (1998) ``very active'' stars,  among which the 
authors find that stellar activity may affect the Str\"{o}mgren colors. Those 
``very  active'' stars have $\log R_{HK}^{\prime} > -4.3$, and tend to be 
either very young, or in close or interacting binary systems. We find that the 
lower limit on X-ray luminosity for their ``very active'' category of stars is 
$\log (L_{\rm X}/L_{\rm bol}) \approx -4.1$. There are
only five stars (out of 146) with such high X-ray luminosity in our 
RASS-BSC-calibration set, 
none of which are within 25 pc of the Sun.
Hence the sample of RASS-BSC stars that we have been using, as selected 
from the compilation of H\"{u}nsch et al. (1998), avoids the uppermost end of 
the stellar X-ray luminosity function.

In summary, we find no evidence for a trend in 
${\rm [Fe/H]_{spec}-[Fe/H]_{phot}}$ with $\log (L_{\rm X}/L_{\rm bol})$, 
regardless of whether ${\rm [Fe/H]_{phot}}$ is based on equation (1) or 
the previous widely-used calibration of Schuster \& Nissen (1989).
We conclude that for most normal single stars, there is no need to apply a
correction for chromospheric activity to metallicity calibrations based on
Str\"{o}mgren photometry.

\newpage
\begin{center}
\large Figure Captions
\end{center}

\figcaption{
Histograms of ${\rm [Fe/H]_{phot} - [Fe/H]_{spec}}$, for the current
calibration (MS04, solid line), ML02 (dotted line), and SN89 (dashed
line), together with Gaussian fits to each.  These histograms provide
a visual representation of how well the various photometric
calibrations reproduce the spectroscopic metallicities.  The
calibrations of MS04 and ML02 are quite similar, but that of SN89 fits
the calibration stars less well.  
\label{fig1}}

\figcaption{
The difference between 
photometric metallicities derived from the current calibration (MS04) and 
that of Martell \& Laughlin (ML02) versus 
spectroscopic metallicity ${\rm [Fe/H]_{spec}}$ for the calibration stars.
\label{fig2}}

\figcaption{
The difference between photometric metallicities derived from
the current calibration (MS04) and the standard calibration of Schuster \& 
Nissen (SN89) versus ${\rm [Fe/H]_{spec}}$ for the MS04 calibration stars.
\label{fig3}}

\figcaption{
Distribution of distances for the RASS-BSC-calibration stars.
\label{fig4}}

\figcaption{
Distribution of ${\rm [Fe/H]_{spec}}$ for the RASS-BSC-calibration stars.
\label{fig5}}

\figcaption{
Distance versus $\log L_{\rm X}$ for the RASS-BSC-calibration stars.  The 
solid line shows the maximum distance at which a star of a given $L_{\rm X}$ 
would have been detected in the \textit{ROSAT} All-Sky Survey.
\label{fig6}}

\figcaption{
Difference between ${\rm [Fe/H]_{spec}}$ and photometric 
metallicity, calculated using the MS04 calibration,
versus $\log (L_{\rm X}/L_{\rm bol})$ for 
RASS-BSC-calibration stars.
\label{fig7}}

\figcaption{
Difference between ${\rm [Fe/H]_{spec}}$ and photometric metallicity, 
calculated using the MS04 calibration, versus $\log (L_{\rm X}/L_{\rm bol})$ 
for RASS-BSC-calibration stars within 25 pc.
\label{fig8}}

\figcaption{
Difference between ${\rm [Fe/H]_{spec}}$ and photometric metallicity, 
calculated using the SN89 calibration, versus $\log (L_{\rm X}/L_{\rm bol})$
for RASS-BSC-calibration stars. 
\label{fig9}}

\figcaption{
Difference between ${\rm [Fe/H]_{spec}}$ and photometric metallicity, 
calculated using the SN89 calibration, versus $\log (L_{\rm X}/L_{\rm bol})$
for RASS-BSC-calibration stars within 25 pc.
\label{fig10}}

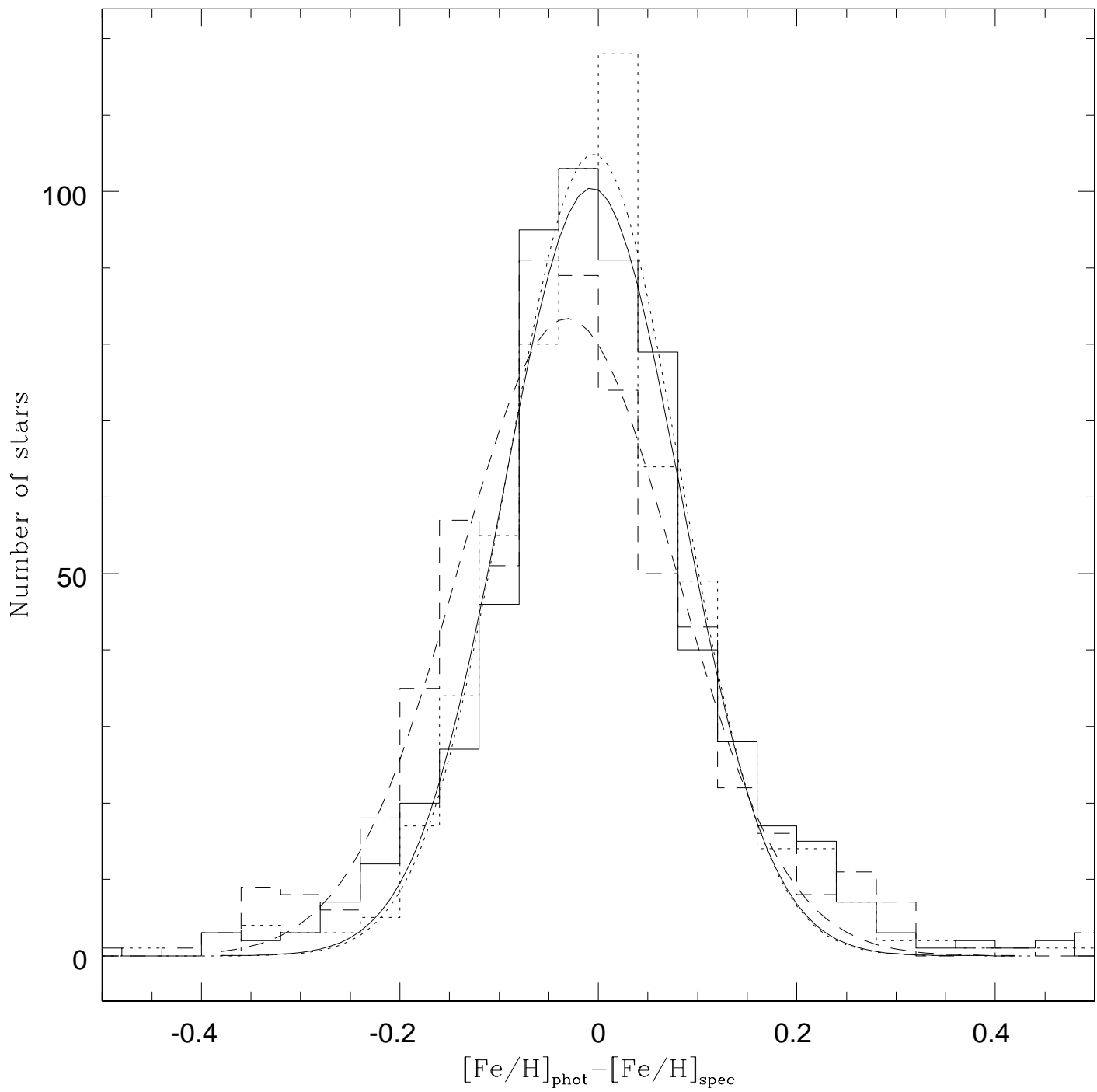
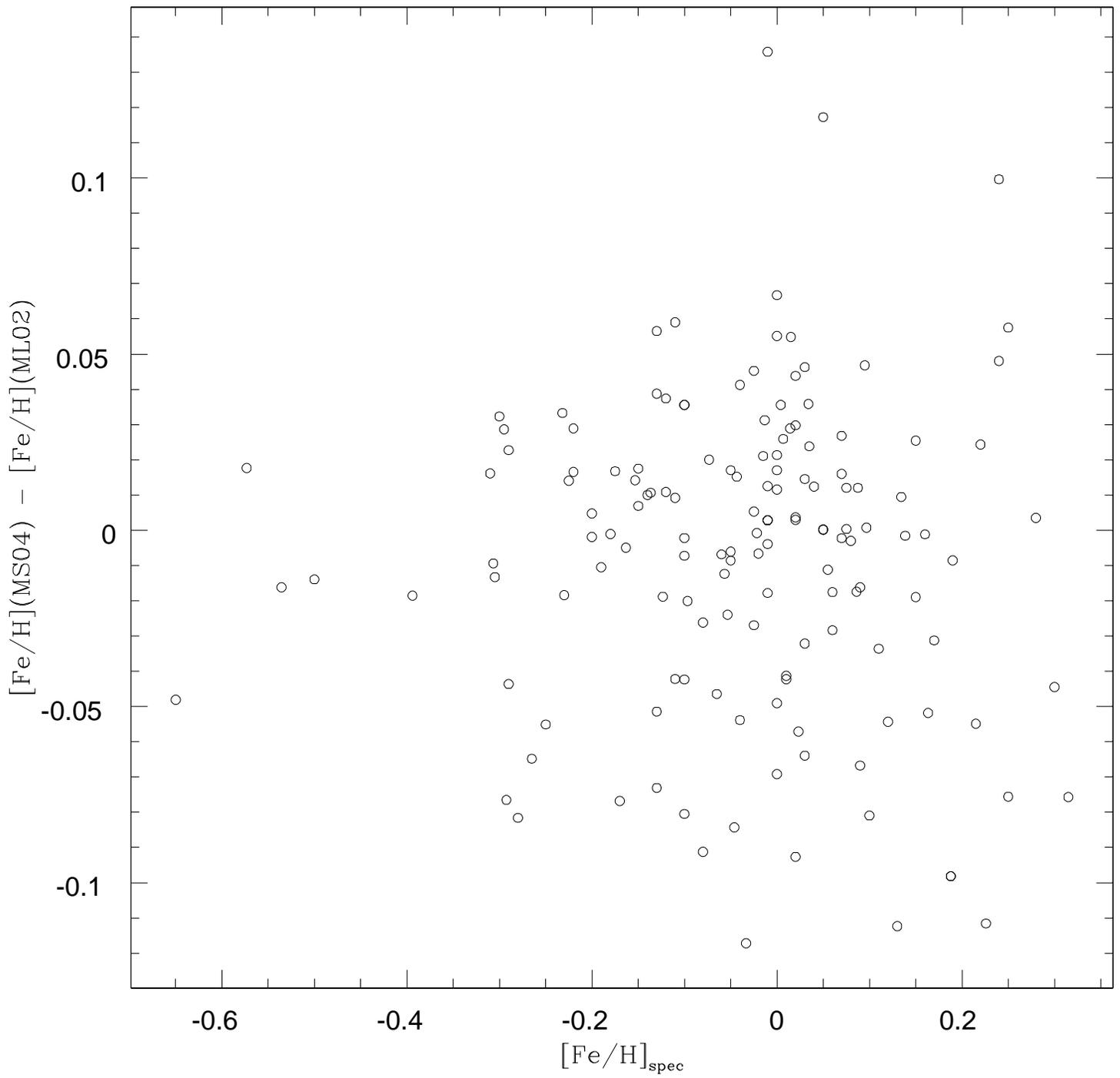
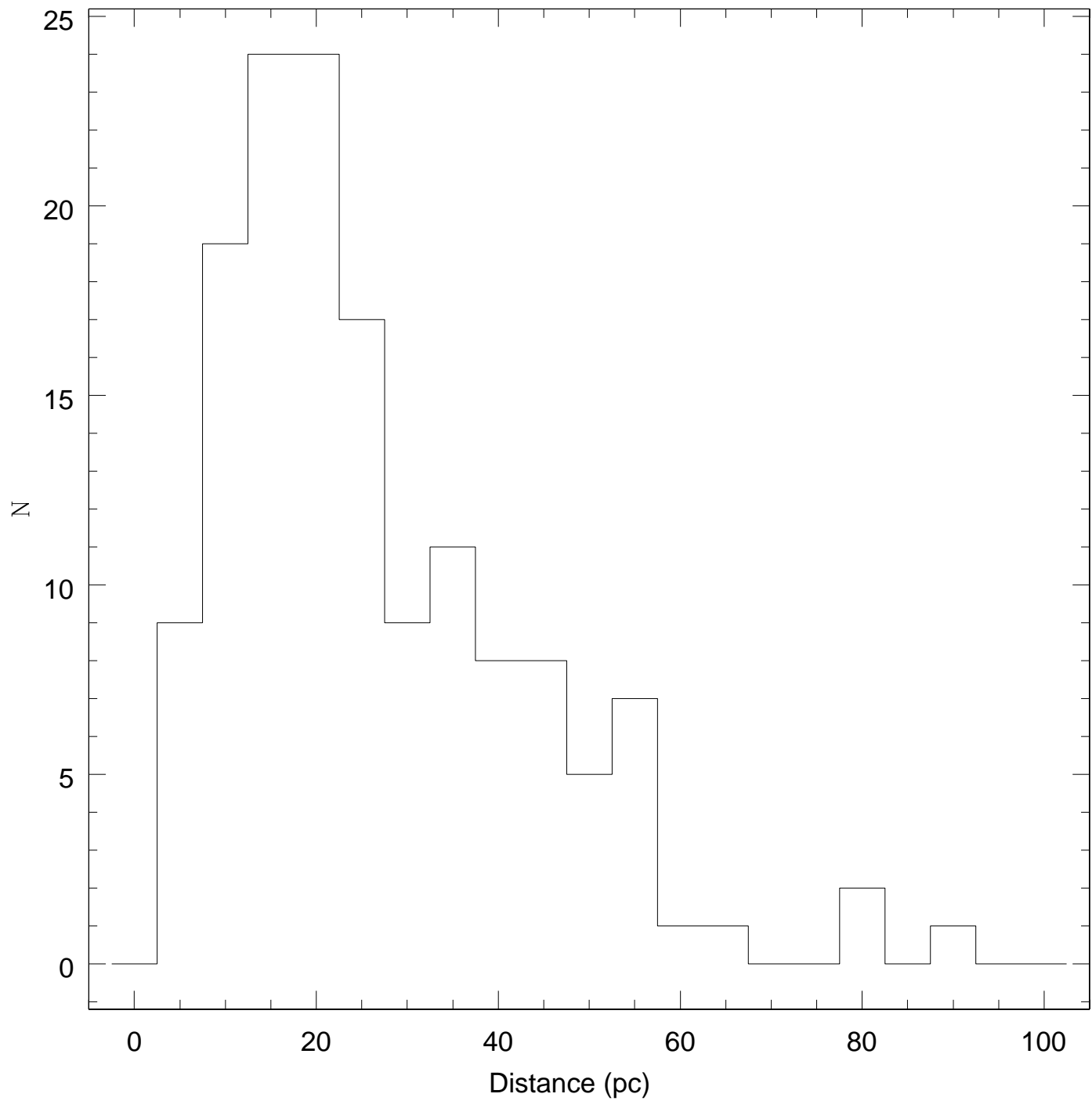
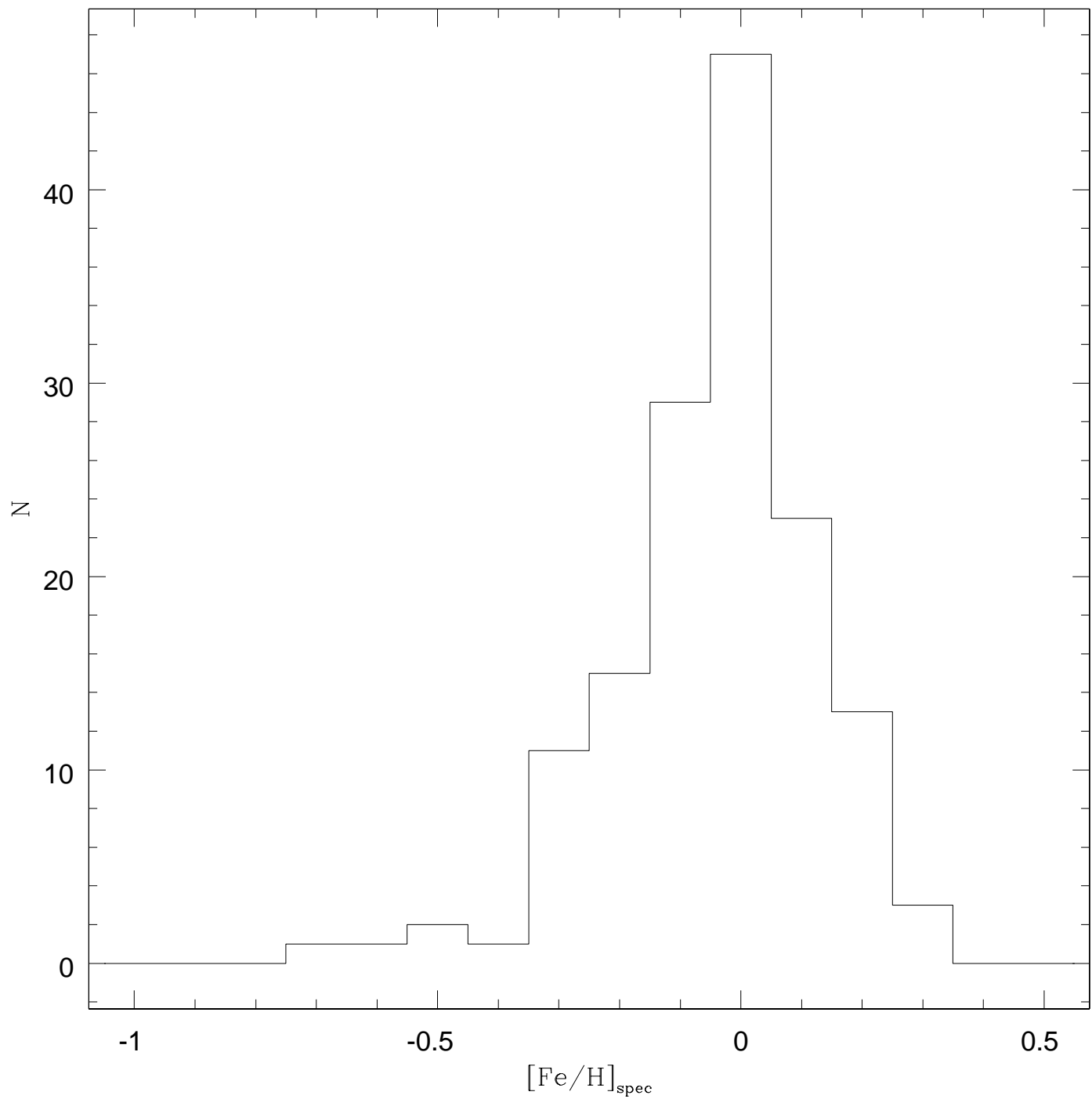
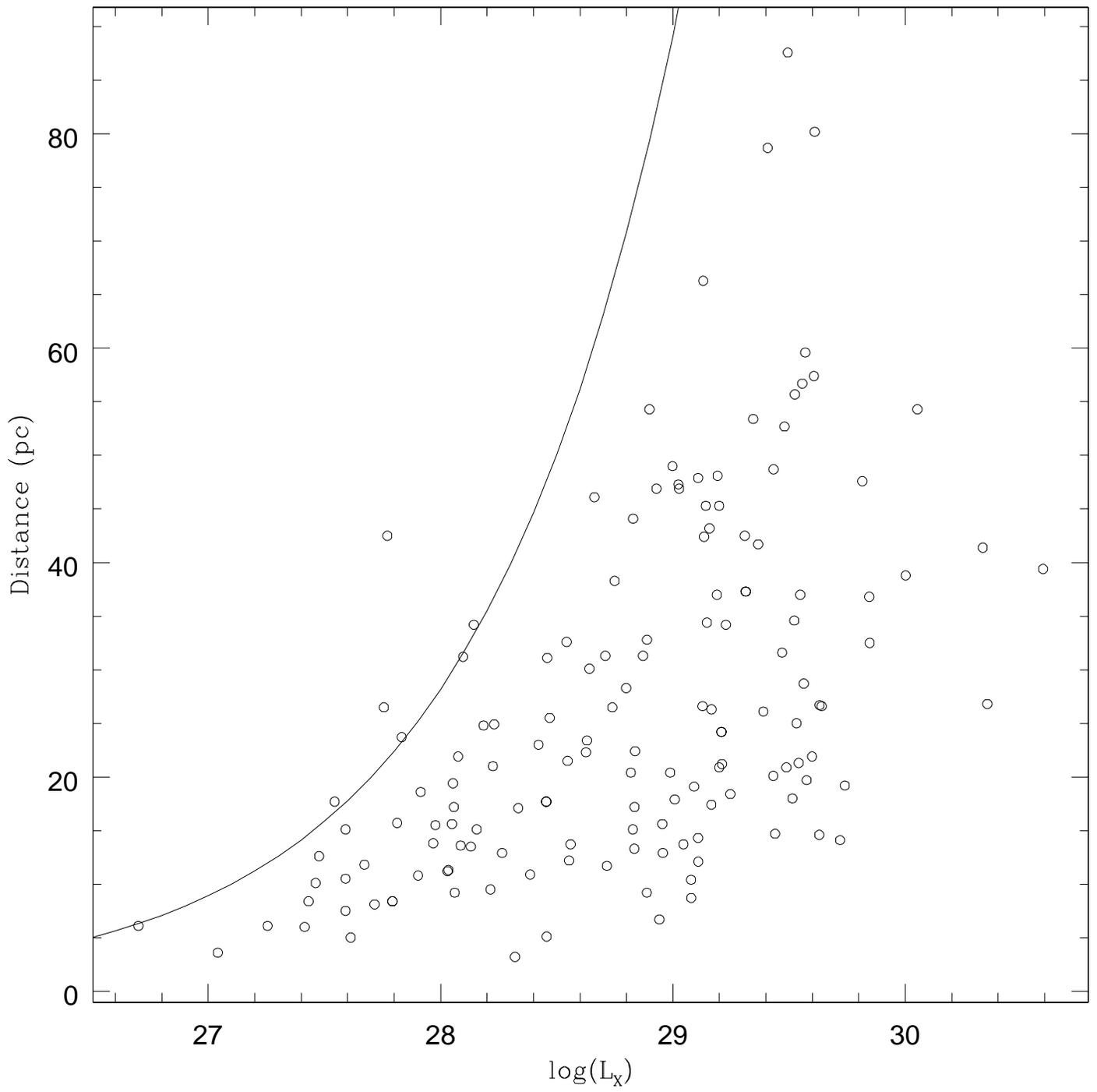
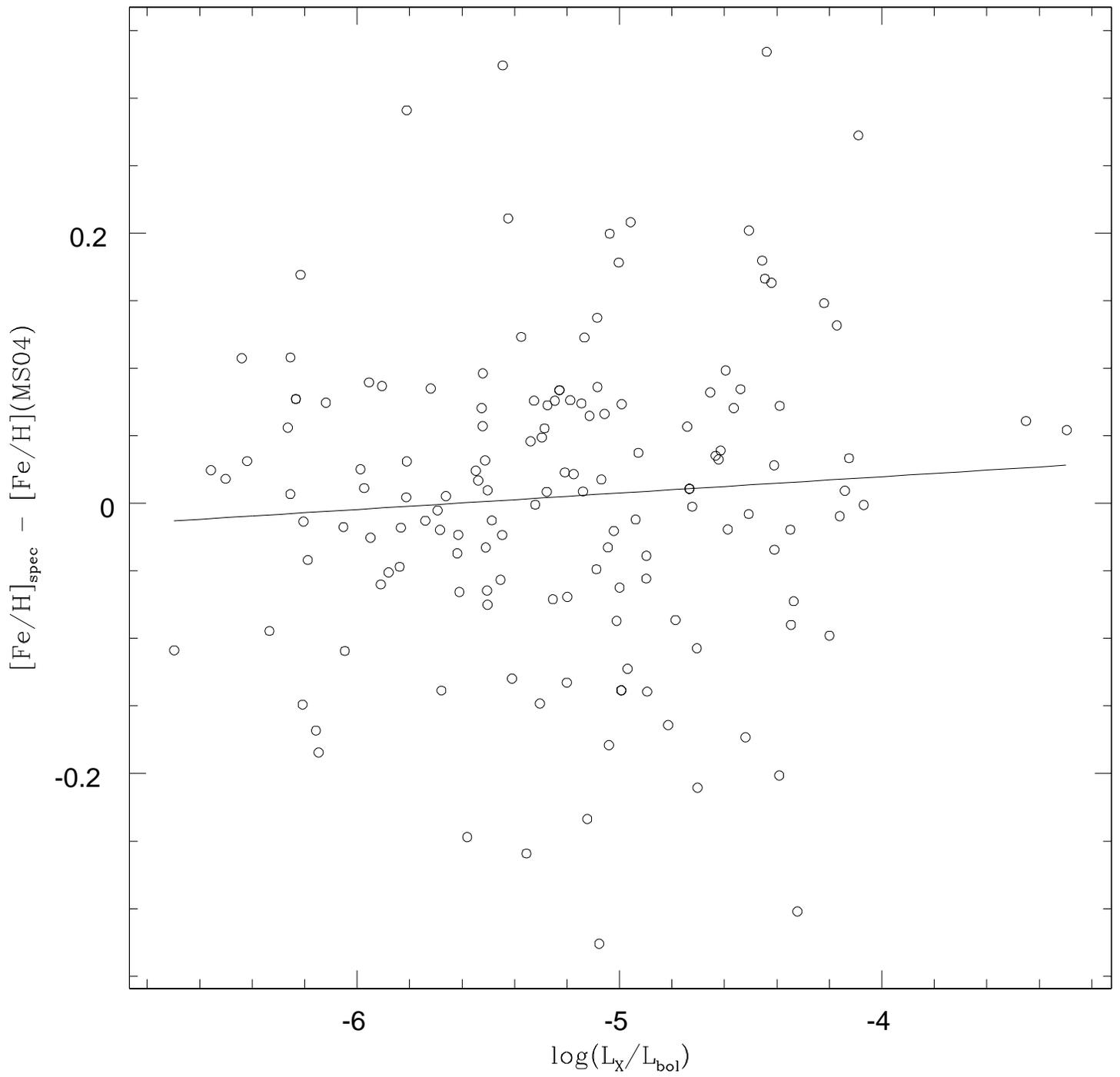
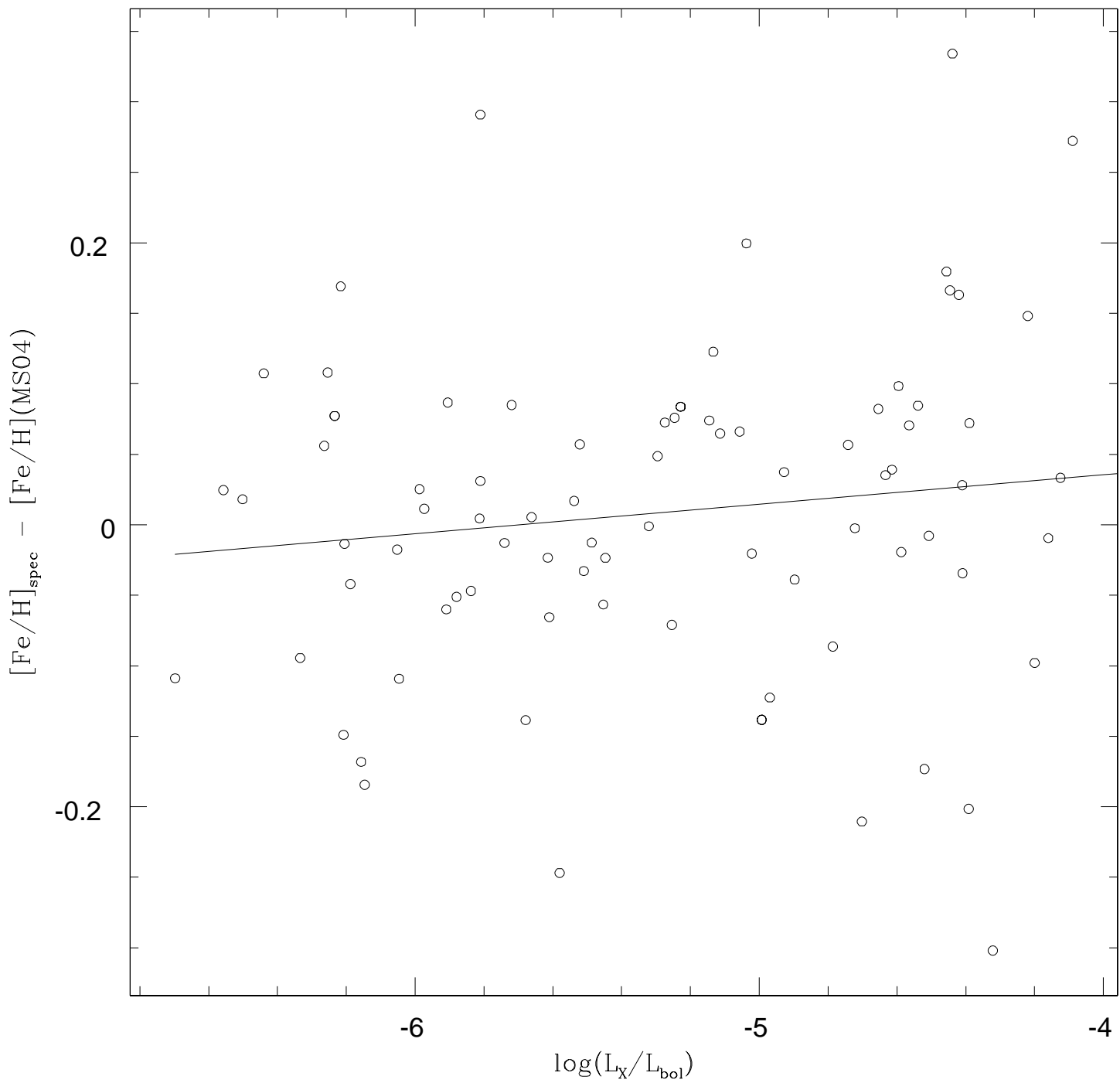
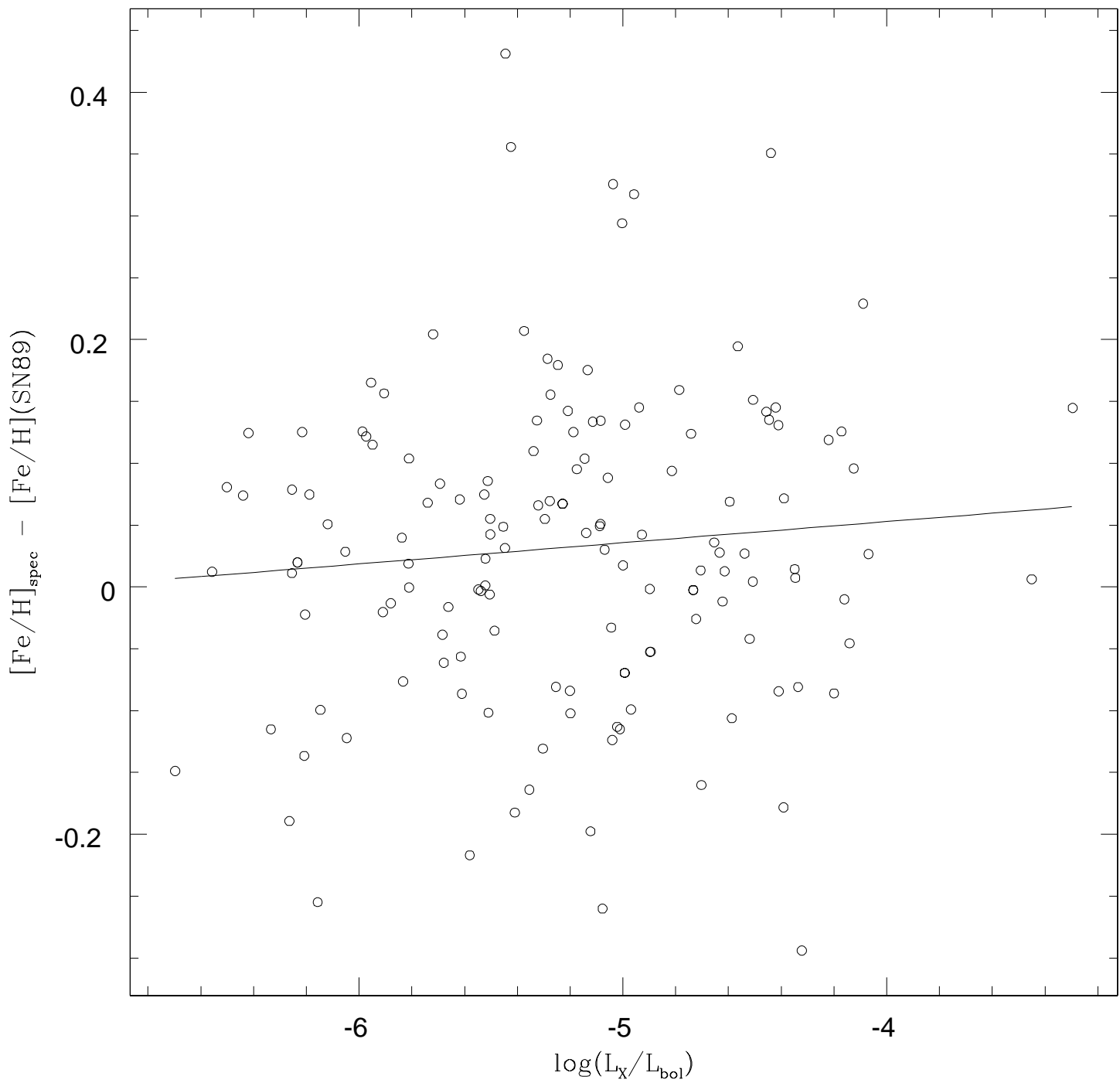
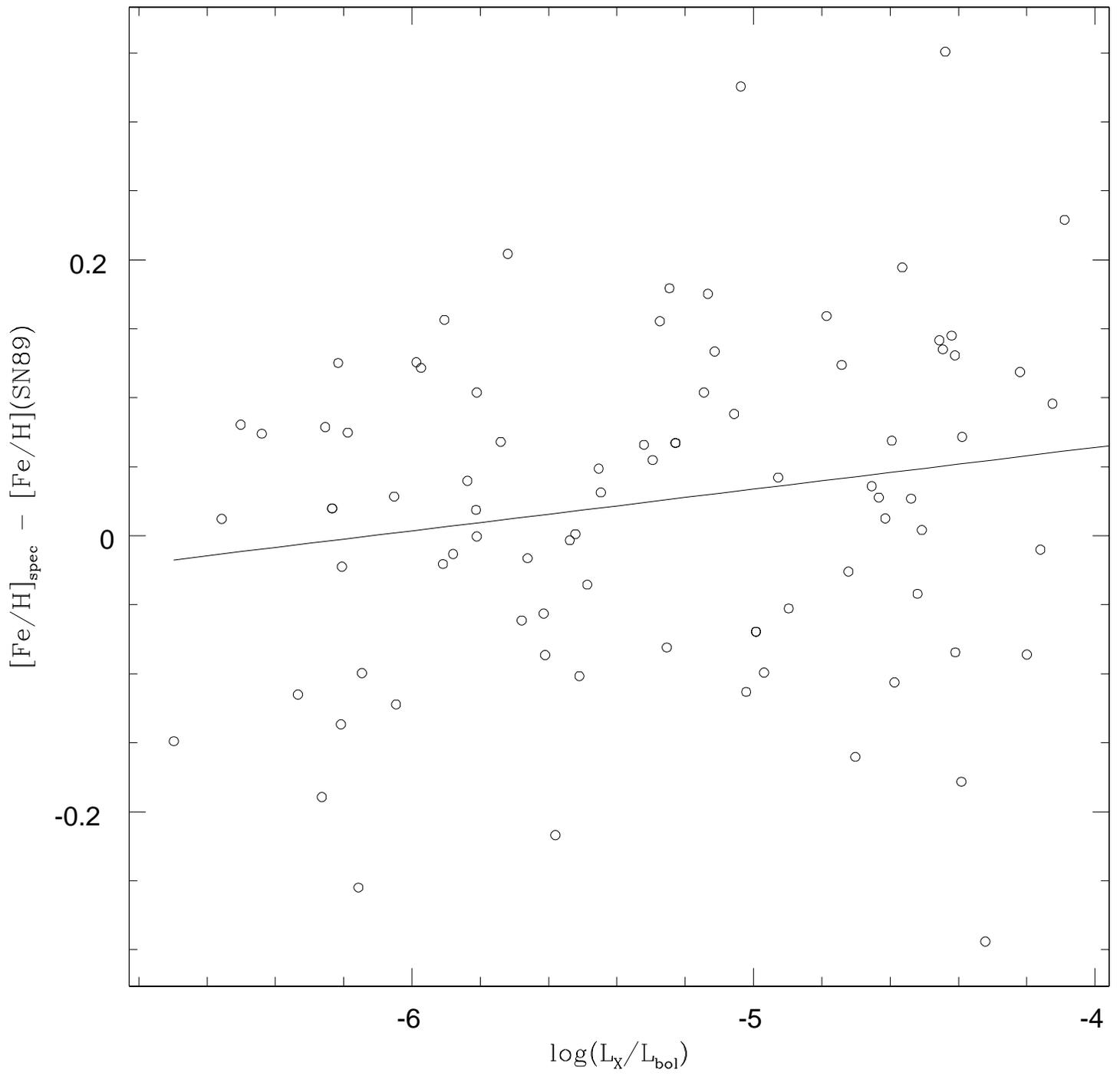
\end{document}